\documentclass[doublecol]{epl2} 
\usepackage{graphicx}

\title{Noise Enhanced Activity in a Complex Network}

\author{Anshul Choudhary, Vivek Kohar \and Sudeshna Sinha}

\institute{ Indian Institute of Science Education and Research (IISER) Mohali, Knowledge City, SAS Nagar, Sector 81, Manauli PO 140 306, Punjab, India}

\abstract{ We consider the influence of local noise on a generalized
  network of populations having positive and negative feedbacks. 
  The population dynamics at the nodes is nonlinear, typically
  chaotic, and allows cessation of activity if the population falls
  below a threshold value. We investigate the global stability of this
  large interactive system, as indicated by the average number of
  nodal populations that manage to remain active. Our central result
  is that the probability of obtaining active nodes in this network is
  significantly enhanced under fluctuations. Further, we find a sharp
  transition in the number of active nodes as noise strength is
  varied, along with clearly evident scaling behaviour near the
  critical noise strength. Lastly, we also observe noise induced
  temporal coherence in the active sub-network, namely, there is an
  enhancement in synchrony among the nodes at an intermediate noise
  strength.}


\begin{document}

\maketitle

\section{Introduction}

Recent years have witnessed a rapidly growing interest in network
research primarily due to its wide applicability in modeling complex
systems. From computer science and mathematics to physical, chemical,
biological and social sciences\cite{network1, network2, may_book}, researchers
are using ideas from network theory to gain understanding of large
interactive dynamical systems.  One of the classic approaches in this
direction was the work by May where he analyzed the problem of the
stability of a generalized ecosystem by considering a random network
of different species. He proved the influential result that, as the
network becomes sufficiently complex, the system gets unstable and the
probability of survival becomes vanishingly small \cite{may72}. May
assumed an a priori equilibrium state of the system and used local
stability analysis to obtain his results. More recently, persistent
activity, and the size of the asymptotic active sub-network, have been
investigated as indicators of the global stability of a model
ecosystem \cite{sinha}, and this criterion also yielded the same
dependence on the complexity as May's analysis.

In another direction, in recent years, several studies have been
reported on the effect of noise in nonlinear systems.  New
counterintuitive phenomena emerging from the interplay of noise and
nonlinearity, such as stochastic resonance \cite{sr}, noise enhanced
stability \cite{nes} and noise delayed extinction \cite{nint} have
been observed.  Motivated by these, we revisit the important problem
global stability of complex systems to include the {\em role of
  noise}, since noise is ubiquitous in such systems. Our principal
question is the following: do fluctuations allow a larger, or smaller,
number of active populations to exist in a complex web? Namely, can
stochastic influences actually yield a larger number of actives nodes
in the eco-network, on an average? Such questions are relevant, in
general, to complex networks subject to fluctuations, where the
evolution of the state of the nodes is nonlinear and allows cessation
of activity.


\section{Deterministic Network Model}

In population dynamics, a generic ecosystem can be modeled as a
complex network, where each node represents a population and the
interaction between the nodes are modeled through links which
determine the strength and nature of the mutual interaction. Namely,
one considers an ensemble of distributed populations which interact
according to predefined rules, which account for various types of
interactions, such as mutualism, predator-prey, competition, etc.
  
Specifically, here we consider $N$ populations evolving in a complex
web, where the nodal population dynamics is represented by a local
nonlinear map $f$, and the interactions are given in most general
terms by an interaction matrix or {\em Community Matrix} {\bf J}
\cite{levin}, whose elements {\em $J_{ij}$ } represent the effect of
species {\em j} on species {\em i}.  The nature and strength of
interaction between node $i$ and node $j$ is given by the sign and
magnitude of the element ($J_{ij}$) in the interaction matrix ${\bf
  J}$. We consider the most general case where the coefficients can be
asymmetric ($J_{ij} \neq J_{ji}$) and can be either positive or
negative.

So the dynamical state of each node \emph{i} ($i=1,....,N$) at time
(or generation) $n$ is denoted by $x_{i}(n)$, which represents the
scaled $i^{th}$ population, and its time evolution under the
prototypical Lotka-Volterra type interaction \cite{ sinha, LV_ref, robust} is given by
\begin{equation}
x_{i}(n+1) = f\left[{x}_{i}(n)\left(1 + \sum_{j}J_{ij}x_{j}(n)\right)\right] 
\label{det_eqn}
\end{equation}
where $f$ represents the local on-site dynamics, and connectivity
matrix ${\bf J}$ represents the positive and negative feedback amongst
the nodal populations.

In this work we choose a prototypical map $f$, modeling population
growth of species with non-overlapping generations, given by a
modified Ricker (Exponential) Map as follows :
\begin{equation}
f(x) = \left\{ \begin{array}{rl}
x e^{r(1-x)} & \ \ \ \ \mbox{ if $x > x_{threshold}$,} \\
0 & \ \ \ \ \mbox{ otherwise,}
\end{array} \right.
\label{map}
\end{equation}
where growth rate $r$ is the nonlinearity parameter yielding behaviour
that ranges from fixed points and periodic cycles to chaos. Here
$x_{thresold}$ is a threshold value, typically very small ($<< 1$),
giving the minimum population density necessary for any further
activity to occur. Namely when the population density falls below this
level, there is extinction \cite{footnote}.  Further note that very
large population density is also detrimental, as the population
dynamics maps large $x$ back to the extinction zone, leading to
inactivity in subsequent generations.

The connectivity matrix {\bf J} in Eqn. \ref{det_eqn} is a matrix
where an element is non-zero with probability $C$ ($0 \le C \le 1$),
i.e. $C$ represents the connectivity of the system. The diagonal
entries $J_{ii}=0$, and this indicates that in the absence of
interactions the local nonlinear map (Eq.~(\ref{map})) completely
determines the dynamical state of each node. The nonzero entries in
the matrix are chosen from a normal distribution with mean $0$ and
variance $\sigma^{2}$.

Also note that the local dynamical map belongs to the class of maps
defined over the semi-infinite interval $[0,\infty)$, rather than a
finite, bounded interval (such as the commonly used logistic
map). This allows us to explore arbitrary distributions of coupling
between nodes, unlike maps bounded in an interval, which are well
behaved only for restrictive coupling schemes.

Now, consider the initial states of all the $N$ nodes to be randomly
distributed about $x=1$. In the course of evolution of the network, if
a nodal population below certain minimum sustainable population $x_{threshold}$, 
the node stops being active and
subsequently has no interaction with the rest of the network. It was
observed in \cite{sinha, robust}, that as a result of interactions, the number
of active nodes (i.e. nodes $i$ with $x_i > x_{threshold}$) decreased rapidly and
eventually attained a steady state \cite{sinha_density}.
This can be rationalized by the fact that in the initial stages the
population at each node undergoes strong fluctuations due to
interactions with other nodes coupled to it, resulting in the
cessation of the activity of a large number of nodes\cite{stat_extinct}. Within a very
short time, the effective number of interacting nodes decreases due to
extinction and consequently the intensity of such fluctuations is also
reduced. The number of active nodes in the asymptotic steady state,
denoted as $N_{active}$, was found to be independent of $N$
\cite{sinha}. So, the evolution of this complex network leads to an
active sub-network whose size does not scale with the size of the
initial network, i.e. the asymptotic state is characterized by a
macroscopic quantity that is non-extensive. The nonextensivity for the
active subnetwork has significant implications, as it indicates that
there exists a characteristic, rather small, size for a globally
stable web of populations \cite{wilmers}.

\section{Stochastic Nodal Dynamics in the Network}

Now, most real ecosystems cannot be modeled by deterministic networks
alone, as noise is ubiquitous. The stochasticity in models are
necessitated by the fact that habitats are typically open systems
subject to external influences such as migrations. Furthermore there
are fluctuations in population size due to random demographic
events\cite{demographic_stochasticity}. So it is of considerable interest 
to ascertain if the emergent
active sub-networks found in deterministic systems are robust against
small perturbation or noise. {\em Further, it would be very
  interesting to ascertain if noise aids, or hinders, the average
  activity of the web.}

In order to investigate these questions we study the effects of
stochasticity in the network above, by considering the evolution
equation (eq.~(\ref{det_eqn})) under a random {\em additive noise}
$\xi(t)$. Here $\xi(t)$ is a gaussian white noise with zero mean and
correlation function given by $\langle \xi_{i}(t)\xi_{j}(t^{'})\rangle
= \eta\delta(t-t^{'})\delta_{ij}$ $(i,j=1,2,...,N)$, where $\eta$
governs the strength of noise. Therefore, the evolution of the local
nodal populations in the network is now governed by the equations:
\begin{equation}
x_{i}(n+1) = f\left[{x}_{i}(n)\left(1 + \sum_{j}J_{ij}x_{j}(n)\right)\right] + \xi_{i}(t) 
\label{stochastic_eqn}
\end{equation}

In order to understand the macroscopic or collective response of the
system to noise at the microscopic level, we calculate the the number
of active nodes (i.e. nodes with $x > x_{threshold}$) in the network averaged over
a long time, and further averaged over different realizations of the
system. We denote this time and ensemble average of the number of
active nodes as $\langle {N}_{active} \rangle$. In this work, we calculate this
quantity for networks with varying local dynamics and connectivities,
under a wide range of noise strengths.

\bigskip
    
{\em Effect of Nodal Noise on the Activity of the Network:}\\

\begin{figure}
\onefigure[scale=0.35,angle = 270]{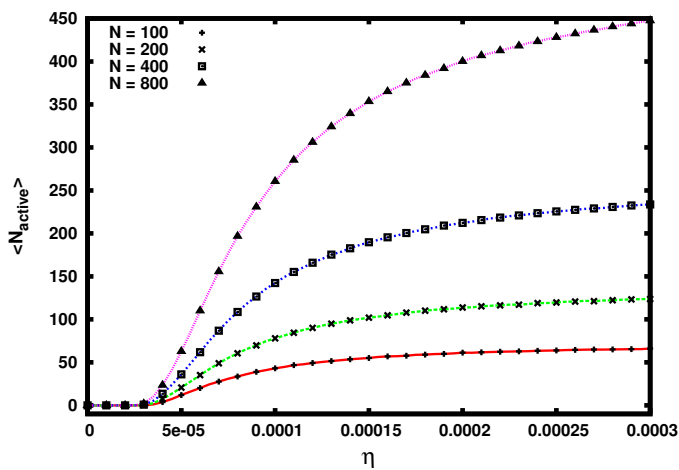}
\caption{Variation in the average number of active nodes $\langle
  {N}_{active} \rangle$ (see text), with respect to increasing noise
  strength, indicating a sharp rise in the average number of active
  nodes after a critical noise strength, $\eta_{c}$. System sizes are
  $100, 200, 400$ and $800$ and threshold population size
  (cf. Eqn. \ref{map}), $x_{threshold}=0.0001$. Here the network
  connectedness, $C=1$, mean interaction strength is $0$ and standard deviation,
  $\sigma=0.1$. }
\label{unscaled}
\end{figure}

Fig. \ref{unscaled} shows how the average number of active nodes
varies as the strength of noise at the nodes increases. It is clearly
evident that there is very sharp transition at noise strengths $\sim
\eta_{c}$. Below this critical noise strength (i.e. $\eta<\eta_{c}$),
noise has no discernable effect on network activity, while above
$\eta_{c}$ there is a {\em very significant jump in the number of
  active nodes in the network}. In the noise regime much larger than
$\eta_{c}$, the system settles to a non-equilibrium steady where a
constant number of active nodes is maintained on an average. Namely,
at high enough noise strengths $\eta >> \eta_{c}$, the mean number of
active nodes saturates to an asymptotic value. So clearly there is a
\emph{noise induced transition}, from a system where the emergent
active sub-network is very small to a system with a large active
sub-network.

\begin{figure}
\onefigure[scale=0.35,angle = 270]{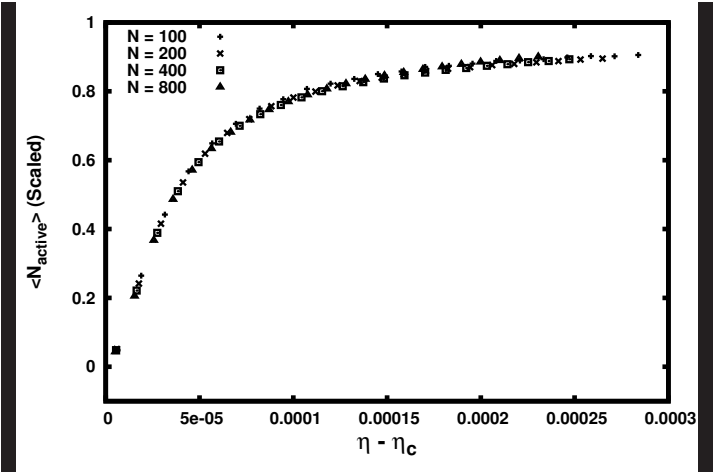}
\caption{ Scaling function for which the relation between the average
  number of active nodes, $\langle N_{active} \rangle$, and noise
  strength collapse on a single curve for all system sizes, $N$. (See
  Eqn. \ref{scaling_fn}.)}
\label{scaled}
\end{figure}

In order to account for the effect of network size on the nature of the
transition, we have done the finite size scaling for a range of system
sizes. As evident from Fig \ref{scaled}, the transition curves
for all system sizes collapse to a single curve, allowing us to
calculate the value of the critical noise strength $\eta_{c}$ in the
thermodynamic limit. The scaling function near the critical region is
given by: 
\begin{equation}
P \sim N^{\alpha / \beta}\Theta(N^{-1 / \beta},(\eta-\eta_{c})) 
\label{scaling_fn}
\end{equation}
where, $\eta_{c} = 0.00003\pm0.000001$, $\alpha / \beta = 0.1\pm0.02$
and $1 / \beta = 0.94 \pm0.005$. Note that the critical noise strength
$\eta_c$ is significantly smaller than the threshold of survival
$x_{threshold}$ of the individual populations.

Also notice that the number of active nodes saturates to a steady mean
size for high noise strengths. However, in contrast to the
deterministic network, we now observe that the size of the asymptotic
active sub-network depends on $N$. Namely, under reasonable noise, the
network settles down to an active sub-network that scales with the
size of the original network, unlike the deterministic case where the
network settled down rapidly to an active sub-network of small
characteristic size, for all system sizes.

To further understand the behavior of the population dynamics at a
node under the influence of noise, we analyze this problem as a system
exploring a state space with a zone of inactivity, bounded by
$x_{threshold}$. Now in the deterministic model
(Eqn.~(\ref{det_eqn})), when a population enters the inactive zone, it
cannot leave the region. So the extinction threshold density
$x_{threshold}$ acts as an absorbing boundary condition, leading to
large-scale extinctions and very small persisting active
sub-networks. However in the presence of noise, populations can be
pushed out of this zone, i.e. noise allows populations to escape the
inactive region and revive to become active again.

\begin{figure}
\onefigure[scale=0.35,angle = 270]{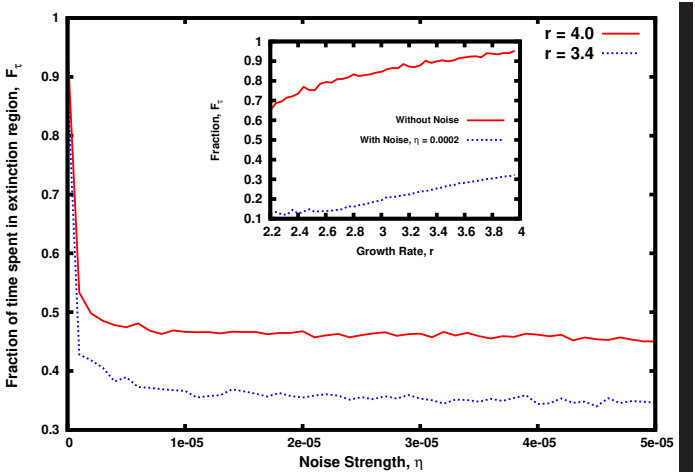}
\caption{Variation of the fraction of time spent by the system in
  the inactive region $[0:x_{threshold}]$ with respect to noise strength
  $\eta$ (main) and with respect to growth rate $r$ in the local
  population dynamics given by Eqn. \ref{map} (inset). Here the
  network consists of $100$ nodes, subjected to gaussian white noise, and we
  leave a transience of $10000$ steps.}
\label{fraction}
\end{figure}

One can calculate the fraction of time spent by the system in the
inactive region $([0:x_{threshold}])$, {\em $F_{\tau}$}, in the
deterministic case, and in the presence, of noise. Representative
results are displayed in Fig. \ref{fraction} and these suggest that as
noise strength varies there is a very sharp fall in {\em $F_{\tau}$}.
The inset shows the variation of {\em $F_{\tau}$} with growth parameter
{\em r} of the local dynamics, in the presence and in the absence of
noise. The results suggest that the noise-free system spends most of
the time in the inactive zone, as once the node goes inactive, it
remains inactive due to the nature of the predator-prey type
interaction. On the other hand, when there are external fluctuations,
the system spends a significantly less amount of time in the
inactive region, which results in much enhanced average activity.

Also note that in the particular example shown in inset of
Fig. \ref{fraction}, with noise strength, $\eta = 0.002$, the fraction
of time spent in the extinction zone is about 30 \%, for growth rate
parameter $r=4$ in the local population dynamics. This implies that
about 70 \% of the nodes in the network are active, on an average, in
the saturation limit. This estimate agrees very well, quantitatively,
with results from simulations shown in Fig. \ref{unscaled}.

\section{Influence of a sub-set of noisy nodes on network activity}

Now, all the nodes in the network are not necessarily influenced by
noise at all instants of time. In fact, at any given time there may be
only a certain fraction of noisy nodes, and this fraction too may
change with time depending on global influences. In this section, we
consider a network where a different set of random nodes are
influenced by noise at every instant of time, with the total number of
noisy nodes being fixed. That is, dynamically, a certain fraction of
the network is subject to fluctuations.

\begin{figure}
\onefigure[scale=0.35,angle = 270]{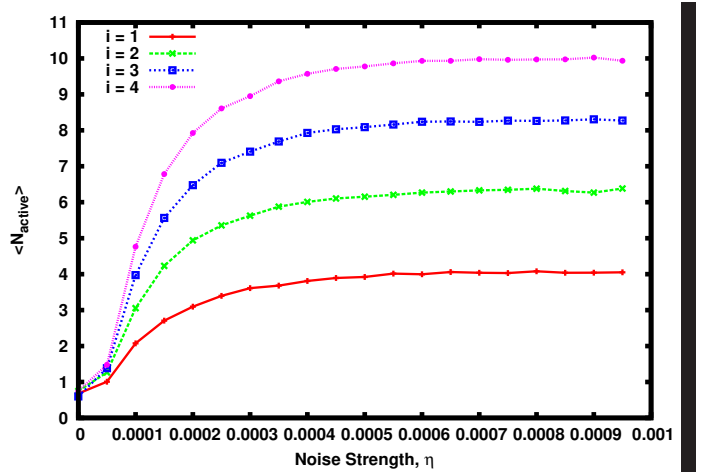}
\caption{Variation of the time averaged number of active nodes in the
  network of size $100$, with respect to noise strength, with the
  number of nodes subject to noise being: (a) $1$ (b) $2$ (c) $3$ and
  (d) $4$. Note that $\langle N_{active} \rangle$ saturates to some
  maximum asymptotic value with increasing noise strength.}

\label{noisy_nodes}
\end{figure}

Fig.~\ref{noisy_nodes} shows the transition from a small number of
active nodes to a high number of active nodes, as the noise strength
increases, for a system with different number of nodes subject to
fluctuations. It is clear that the average number of active nodes
saturates to a high mean value, at noise strengths larger than
$\eta_c$, with the maximum $\langle N_{active} \rangle$ depending on
the fraction of noisy nodes in the system.

The interesting question here is the following: {\em What fraction of
  the network needs to be subject to noise in order to obtain the
  noise induced transition to high activity?}  It is evident from
Fig. \ref{noisy_nodes}, that even when just a {\em single} node in the
network is subject to noise, the activity of the network is
significantly enhanced. 

\begin{figure}
\onefigure[scale=0.35,angle = 270]{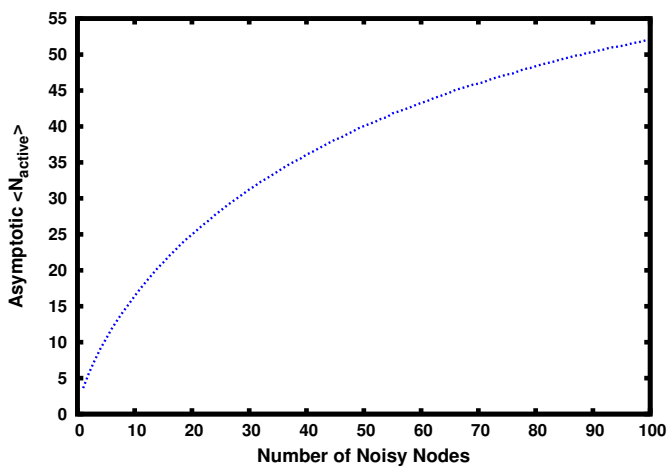}
\caption{Asymptotic $\langle N_{active} \rangle$ in the network of size $100$ (see
  Fig. \ref{noisy_nodes}), as a function of the number of nodes
  subject to gaussian white noise where $x_{threshold}=0.0001$.}
\label{saturation}
\end{figure}

The dependence of the asymptotic $\langle N_{active} \rangle$ on the
fraction of noisy nodes in the network is displayed in
Fig. \ref{saturation}, for representative cases. It is clear from the
figure that the asymptotic value of the average number of active nodes
in the network rises with the increase in number of noisy nodes.



\begin{figure}
\onefigure[scale=0.35,angle = 270]{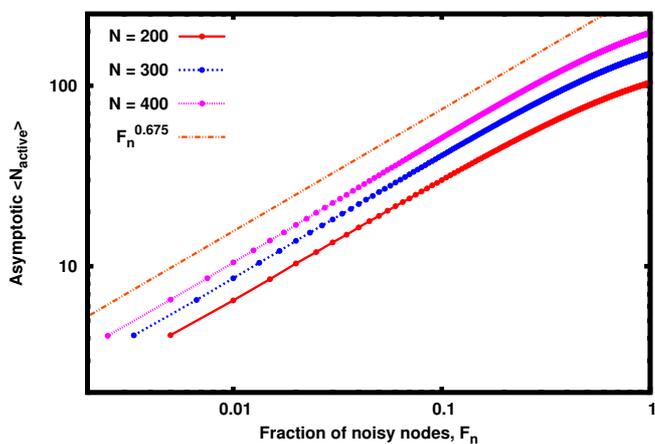}
\caption{Asymptotic ${\langle N_{active} \rangle}$ for different
  network of sizes, as a function of the fraction of nodes subject to
  gaussian white noise where $x_{threshold}=0.0001$.}
\label{saturation_log}
\end{figure}

In order to find the functional dependence of asymptotic value of
$\langle N_{active} \rangle$ on fraction of noisy nodes, 
  $F_{n}$, we study the behavior for different network sizes as shown
in Fig. \ref{saturation_log}. It is evident that the asymptotic
(maximum) value of $\langle N_{active} \rangle$ varies as :
\begin{equation}
\langle N_{active} \rangle \sim {F_{n}^{\alpha}}
\end{equation}
where, $\alpha=0.675$. 

\section{Synchronization}

We now investigate the nature of the {\em local population dynamics}
under the influence of noise. In particular, we study how noise
affects the {\em synchrony of the nodal dynamics}. In order to quantify the
degree of synchronization, we compute an average error function as the
synchronization order parameter, $Z_{sync}$, defined as the mean
square deviation of the instantaneous states of the nodes
\begin{equation}
Z =\frac{1}{N}\sum_{i=1}^N\{[x_i (t) - \langle x (t) \rangle]^2\}
\label{sync}
\end{equation}
where, $\langle x (t)\rangle$ is the space average at time $t$. This
quantity $Z$, averaged over time $n$ and over different initial
conditions, is denoted as $Z_{sync}$.

\begin{figure}
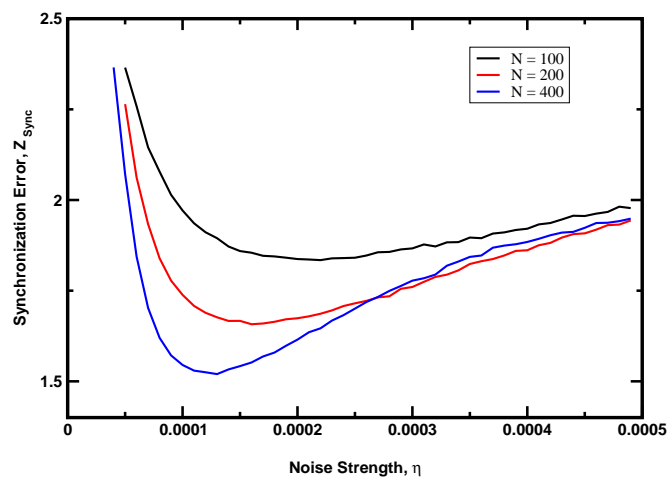

\onefigure[scale=0.35]{sync_grace.eps}
\caption{Synchronization error as a function of noise strength, for three
  different system sizes. Here, gaussian white noise is applied. }
\label{sync_noise}
\end{figure}

We observe an interesting feature in the behavior of synchronization
as noise strength is varied: For a particular intermediate value of
noise strength, there is a significant fall in the synchronization
error. This increase in synchrony is more pronounced for larger
network sizes. So clearly, there is {\em noise induced temporal
  coherence} in the active sub-network, at optimal noise strengths\cite{sync_exp}.
Note that, quite unlike the scenario where synchronization is a result
of the constituents being subject to common noise, here the nodes are
driven by uncorrelated noise that varies from node to node.\\
{\it Note:} It may appear that synchronization error diverges as $\eta \to 0$, but this comes
from the definition of $Z_{sync}$ (Eq.$\ref{sync}$) as number of active nodes tend to zero when $\eta \to 0$.

\section{Generality of our results}

In order to check the generality of our results we have investigated
different kinds of noise and different local dynamics. Specifically,
we demonstrated the increase of active nodes in the system under {\em
  uniform noise}, namely the noise at the nodes was drawn from a
uniform distribution in the interval $[-\eta:\eta]$, where $\eta>0$.

\begin{figure}
\onefigure[scale=0.35,angle = 270]{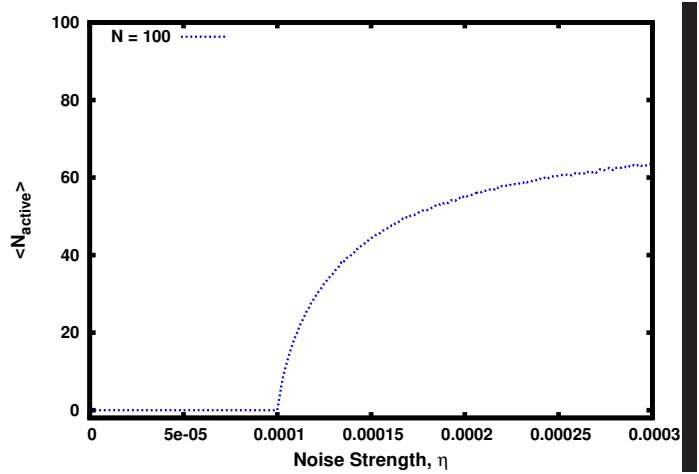}
\caption{Average number of active nodes as a function of $\eta$ (see
  text) for the case of nodes subjected to uniformly distributed
  noise. Here the network size $N =100$ and $x_{threshold}=0.0001$.}
\label{uniform}
\end{figure}

Again we find that the number of active nodes in the network increases
sharply after a critical noise strength. As evident from
Fig.~\ref{uniform}, the critical noise strength for the case of
uniform noise bounded in the interval $[-\eta:\eta]$ is equal to the
extinction threshold, $x_{threshold}$. This is indeed expected, as the
minimum perturbation to push a population out of the inactive zone,
i.e. the minimum noise required to revive, is $x_{threshold}$.

We have also simulated the system for modified logistic growth at the
nodes:
\begin{equation}
f(x) = \left\{ \begin{array}{rl}
r x ( 1 - x ) & \ \ \ \ \mbox{ if $x > x_{threshold}$,} \\
0 & \ \ \ \ \mbox{ otherwise,}
\end{array} \right.
\label{logistic_eqn}
\end{equation}

\begin{figure}
\onefigure[scale=0.35,angle = 270]{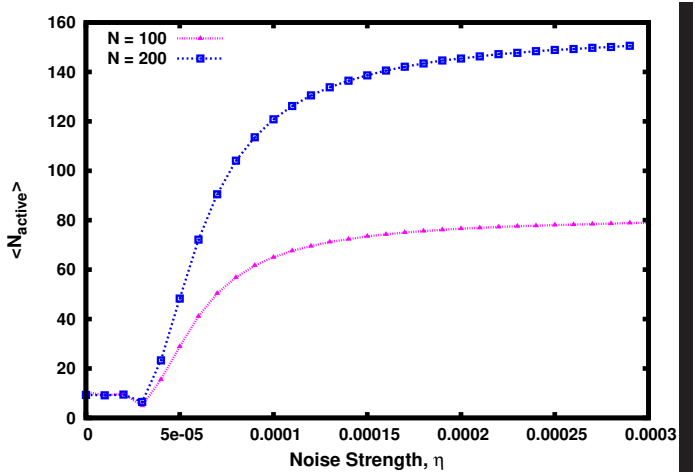}
\caption{Average number of active nodes as a function of $\eta$ (see
  text) for the case of nodes subjected to gaussian white noise. Here
  we have a modified logistic map with $r=4$ (Eqn. \ref{logistic_eqn}) at
  the nodal level and the network size $N =100$, with
  $x_{threshold}=0.0001$.}
\label{logistic}
\end{figure}

The behavior that emerges in this network, under a wide range of
growth rates $r$, is qualitatively similar to that in the network of
Ricker maps discussed above, as evident from Fig.~\ref{logistic}.

Further, we investigated heterogenoeus networks, with a range of
growth rates at the nodal level. The central conclusion, namely the
enhancement of activity in networks under stochastic influence, holds
in such systems as well.


\section{Analysis}

In this section we attempt to gauge the underlying reasons determining
the nature of the noise-induced transition. Specifically, we consider
the case of logistic growth as shown in Fig.~\ref{uniform} where the
system evolves according to
Eqn.[\ref{stochastic_eqn},\ref{logistic_eqn}].  Now, at noise
strengths $\eta > x_{threshold}$, there will be some nodes that will
get pushed out of the extinction region due to noise, while some nodes
will get mapped back to the extinction region at the subsequent
update. So here we will try to get a broad idea of the number of
active nodes in the system at some particular noise strength
$\eta$. Now the fraction of inactive nodes is given by:
\begin{eqnarray}
\frac{N_{inactive}}{N}& \sim & P(\eta<x_{threshold})\int_0^{x_{threshold}} \! \rho^{*}(x) \, \mathrm{d}x \nonumber \\
  &  +~ &P(\eta<-x_{threshold})\int_{x_{threshold}}^{\eta} \! \rho^{*}(x) \, \mathrm{d}x \nonumber \\
  & + &\int_{x^{'}}^{1} \! \rho^{*}(x) \, \mathrm{d}x \hfill
  \label{ndead}
\end{eqnarray}
where, $\rho{*}$ is the invariant measure of the coupled system and
$P(\eta<x_{threshold})$ is the probability of noise $\eta$ being less
than the threshold population density. The first term in
Eqn.~\ref{ndead} corresponds to the fraction of nodes unable to get
out of the extinction region due to sub-threshold noise strength.  The
second term corresponds to the fraction of nodes which were initially
outside the extinction region, but got absorbed into the extinction
region on application of noise. The last term represents the fraction
of nodes that map onto the extinction region due to the nature of the
local dynamics($x^{'}$ represents the inverse image of the logistic map corresponding to $x_{threshold}$). 
Note that noise is drawn from a uniform distribution
bounded in the interval $[-\eta:\eta]$ here. The probabilities for the
noise input to lie in different regions is calculated as below:
\begin{eqnarray}
P(\eta<x_{threshold})&=& \frac{\eta + x_{threshold}}{2\eta}, \nonumber \\
P(\eta>x_{threshold})&=& \frac{\eta - x_{threshold}}{2\eta}, \nonumber \\
P(\eta<-x_{threshold})&=& \frac{x_{threshold} - \eta}{2\eta}
\label{prob}
\end{eqnarray}
Now the calculation of $\rho{*}$ for this high-dimensional coupled
stochastic system is an intractable task, and does not lend itself to
a closed form solutions. So here we approach this problem from a
different perspective. Since we are more interested in the qualitative
behavior of the system, we assume these integrals to be finite
constants whose values are related to values obtained in single
map\cite{invariant_logistic} in some appropriate limit. These
constants are defined below as:
\begin{eqnarray}
\int_0^{x_{threshold}} \! \rho^{*}(x) \, \mathrm{d}x &=& C_{1}, \nonumber \\
\int_{x_{threshold}}^{\eta} \! \rho^{*}(x) \, \mathrm{d}x &=& C_{2}, \nonumber \\
\int_{x^{'}}^{1} \! \rho^{*}(x) \, \mathrm{d}x &=& C_{3}
\label{constant} 
\end{eqnarray} 
Using Eqns. \ref{ndead},\ref{prob} $\&$ \ref{constant}, we get $N_{active} = N - N_{inactive}$ as, 
\begin{equation}
\frac{N_{active}(\eta)}{N} = 1 - (\frac{C_{1}+C_{2}}{2})\frac{x_{threshold}}{\eta} + \frac{C_{2}-C_{1}}{2} + C_{3}
\end{equation}
which can be re-written in a simple form as:
\begin{equation}
\frac{N_{active}(\eta)}{N} =  \alpha - \frac{\beta}{\eta} 
\label{analytic}
\end{equation}
where, $\alpha = 1 + \frac{C_{2}-C_{1}}{2} + C_{3}$ and $\beta = \frac{(C_{1}+C_{2}) x_{threshold}}{2}$. \\
\begin{figure}
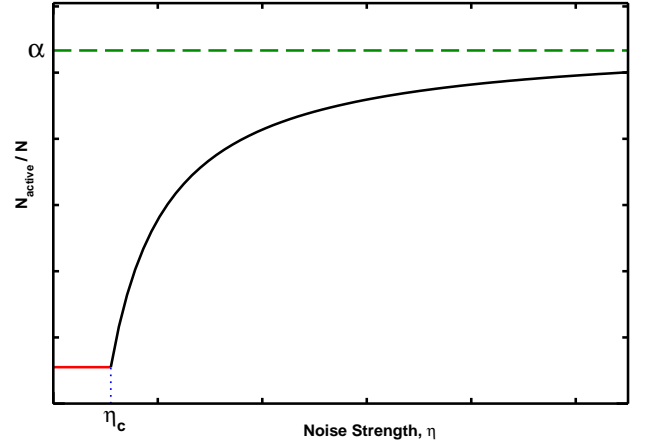

\onefigure[scale=0.35]{analytical.eps}
\caption{Figure shows the plot of Eqn.~\ref{analytic} depicting the
  behavior of fraction of active nodes with respect to noise strength. 
  Parameters $\alpha$, $\beta$ are taken to be $0.1$, $0.0009$ respectively.
  Region $\eta<\eta_{c}$ represents the region where noise has no discernable effect on the 
  activity of the network. Here the noise is drawn from an uniform distribution.}
\label{fig_analytical}
\end{figure}\\
\\
It is evident that our analysis (Fig.~\ref{fig_analytical}) matches
well with the results from numerical simulations (Fig.~\ref{uniform}).

 
\section{Multiplicative Noise}

Lastly, we also studied the evolution of this complex network under
multiplicative noise :
\begin{equation}
x_{i}(n+1) = f\left[{x}_{i}(n)\left(1 + \sum_{j}J_{ij}x_{j}(n)\right) + \xi_{i}(t) \right]  
\label{stochastic_parametric}
\end{equation}
Here $\xi(t)$ is a uniform noise in an interval $[-\eta:\eta]$ with
zero mean and correlation function is given by
$\langle\xi_{i}(t)\xi_{j}(t^{'})\rangle =
\eta\delta(t-t^{'})\delta_{ij}$ $(i,j=1,2,...,N)$, where $\eta$
governs the strength of noise.

\begin{figure}
\onefigure[scale=0.35,angle = 270]{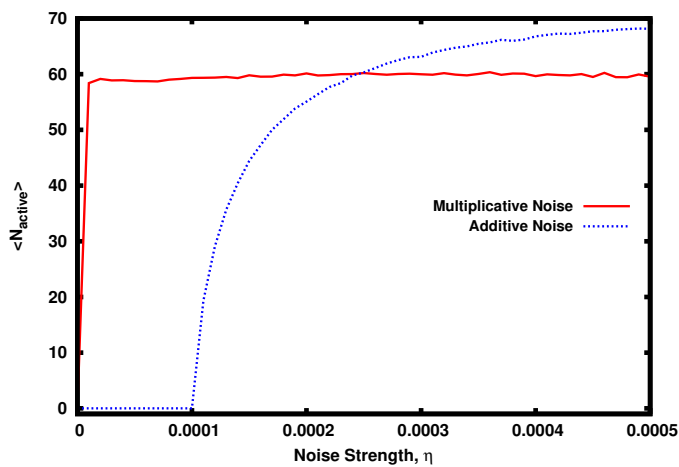}
\caption{Average number of active nodes as a function of noise
  strength. The two curves represent results for multiplicative and
  additive noise. Network size is $100$, activity threshold is
  $x_{threshold} = 0.0001$. Noise here is drawn from an uniform
  distribution in the interval $[-\eta:\eta]$.}
\label{parametric}
\end{figure}

While in the case of additive noise with uniform distribution
(cf. Fig. \ref{uniform}) the critical noise needed for enhancing
activity in the system was exactly the extinction threshold, for the
case of multiplicative noise we observe that the critical noise is far
below $x_{threshold}$. In fact $\eta_c$ tends to zero, as evident from
Fig. \ref{parametric}, indicating the extreme sensitivity of network
activity to multiplicative noise \cite{pf}.

\section{Conclusions}

In summary, we have investigated the role of stochasticity on the
global stability of complex networks. In this study, very general
networks, incorporating positive and negative interactions of the
generalized Lotka-Volterra type, were considered. The population
dynamics at the nodal level was typically chaotic, and allowed
cessation of activity if the population density fell below a certain
threshold value.

Our central result is the following: the probability of obtaining
active nodes in the network is {\em significantly enhanced under
  noise}. Further, we find a sharp transition in the number of active
nodes as noise strength is varied, along with clearly evident scaling
behaviour near the critical noise strength.

\acknowledgments

AC and VK want to acknowledge the financial support from Counsel of Scientific and
Industrial Research(CSIR), India.


\begin{thebibliography}{0}
\bibitem{sinha_density} \Book{The steady state value of $\langle
    N_{active} \rangle$ depends on the choice of $x_{threshold}$.  If
    $x_{threshold}$ is sufficiently high there would be mass
    extinction and $\langle N_{active} \rangle \to 0$ with time. }
\bibitem{invariant_logistic} \Book{The invariant measure for single
    logistic map with $r=4$ has a well known closed form expression
    given by $\rho(x) = \frac{1}{\Pi\sqrt{x(1-x)}}$.}
\end{thebibliography}
\end{document}